\documentclass{PoS}

\pdfoutput=1

\usepackage{enumitem}
\usepackage{url}

\title{APFEL++: A new PDF evolution library in C++}

\ShortTitle{APFEL++}

\author{\speaker{Valerio Bertone}\\
  Department of Physics and Astronomy, VU University, NL-1081 HV
  Amsterdam, \\and Nikhef Theory Group Science Park 105, 1098 XG
  Amsterdam, The Netherlands\\
  E-mail: \email{v.bertone@vu.nl}}

\abstract{I present a preliminary version of {\tt APFEL++}, a C++
  rewriting of the Fortran 77 evolution code {\tt APFEL}. In this
  contribution I discuss the new philosophy adopted for the numerical
  computation of the convolutions, demonstrate the ability to
  reproduce old results in an accurate and fast way, and present an
  original application to the computation of the semi-inclusive
  deep-inelastic-scattering cross section to next-to-leading order in
  QCD.}

\FullConference{XXV International Workshop on Deep-Inelastic Scattering and Related Subjects\\
		3-7 April 2017\\
		University of Birmingham, UK}

\begin{document}

\paragraph{Introduction.}

In the LHC era the need for modern and fast tools for the analysis of
the experimental data is constantly rising. Indeed, meeting the
increasingly high precision of the data entails the growth of the
complexity of the tasks to be carried out to produce accurate
predictions. This often leads existing codes to be continuously
updated with the implementation of new features. This causes an
inevitable stratification that makes the maintenance and the
development of such codes increasingly cumbersome.

Since its born, the {\tt APFEL} code~\cite{Bertone:2013vaa} has
undergone a large number of developments used in many studies (see
\textit{e.g.}
Refs.~\cite{Bertone:2015lqa,Bertone:2016ywq,Ball:2016neh,Bertone:2017ehk,Bertone:2017tyb,Ball:2017nwa,Giuli:2017oii})
and that have led it way beyond the purposes for which it was
originally conceived. In addition, {\tt APFEL} is coded in Fortran
77. This is a perfectly appropriate language for relatively small
programs but lacks of modularity and of tools for an optimal
management of memory and thus it is not suitable for larger
projects. These limitations were compelling reasons to re-think and
re-implement {\tt APFEL} from scratch having in mind the wider range
of applications of the new code.

\paragraph{Basic design of the code.}

The coding language used to write {\tt APFEL++} is C++ as it ensures
modularity, through the definition of objects, and allows for a
dynamic management of memory. We have chosen to use the C++11 official
standard that, as compared to the older major standards, offers a
number of very useful tools and is currently well supported by modern
compilers.

The basic idea behind the structure of the new code is that of
defining a limited number of building blocks that can be used to
construct more complex quantities. The central observation is that, in
the context of collinear factorisation, a relevant quantity $M$ is
typically computed as a Mellin convolution between an
\textit{operator} $O$ and a \textit{distribution} $d$:
\begin{equation}\label{eq:MellinConvolution}
\small
M(x) = \int_0^1dy\int_0^1dz\,O(y)d(z)\delta(x-yz) =
\int_x^1\frac{dy}{y}\,O(y)d\left(\frac{x}{y}\right)=\int_x^1\frac{dz}{z}\,O\left(\frac{x}{z}\right)d(z)\equiv
O(x)\otimes d(x)\,.
\end{equation}
The operator $O$ is typically a process-dependent perturbative
expression that usually contains: ``re\-gu\-lar'' terms whose
integration in Eq.~(\ref{eq:MellinConvolution}) does not present any
complication, ``local'' terms proportional to $\delta(1-x)$, and
``singular'' terms whose singularity in $x = 1$ is subtracted by means
of the so-called plus-prescription. When going to higher perturbative
orders, this kind of expressions may become rather convoluted and make
the numerical integration expensive. The distribution $d$, instead, is
usually a simple function, $e.g.$ a parton distribution or a
fragmentation function, that contains information on the universal
non-perturbative dynamics of some hadron. The function $d$ is
typically fast to access, $e.g.$ through the LHAPDF
interface~\cite{Buckley:2014ana}, and thus is much less expensive than
the operator $O$ in the integral in Eq.~(\ref{eq:MellinConvolution}).

The common strategy to make the computation of integrals such as that
in Eq.~(\ref{eq:MellinConvolution}) fast is that of using
interpolation techniques to precompute the expensive part. More
specifically, the integration variable $z$ is discretised by means of
a \textit{grid} $g\equiv\{z_{0},\dots,z_{N_z}\}$ and the distribution
$d$ is interpolated using a moving set of interpolating functions:
\begin{equation}\label{eq:interpolation}
  d(z) = \sum_{\beta \in g} w_{\beta}(z)d_{\beta}\,,
\end{equation}
where $d_\beta\equiv d(z_{\beta})$ is the value of the distribution
$d$ on the $\beta$-th node of the grid and $w_{\beta}$ is the
interpolating function associated to that node, $e.g.$ a Lagrange
polynomial. Plugging Eq.~(\ref{eq:interpolation}) into
Eq.~(\ref{eq:MellinConvolution}) and assuming the value of $x$ to be
equal to the $\alpha$-th grid node, we find:
\begin{equation}\label{eq:NumConvolution}
M_\alpha\equiv M(x_\alpha) = O_{\alpha\beta}
d_\beta\quad\mbox{with}\quad O_{\alpha\beta}\equiv \int_{x_\alpha}^1
\frac{dy}{y} O(y) w_\beta\left(\frac{x_\alpha}{y}\right)\,,
\end{equation}
where a sum over repeated greek indices is now understood. The value
of $M$ for an arbitrary value of $x$ can then be obtained through
interpolation using Eq.~(\ref{eq:interpolation}). Once the integral
in the r.h.s. of Eq.~(\ref{eq:NumConvolution}) has been computed, the
convolution just amounts of a multiplication between a matrix and a
vector. This operation is very fast and allows one to recompute the
function $M$ with different distributions $d$ very quickly.

We are now able to identify the building blocks we were looking for:
\begin{itemize}[noitemsep]
\item the \textit{grid} $g$, that is the set of nodes over a relevant
  range in the integration variable and of the interpolating functions
  associated to each node,
\item the \textit{distribution} $\{d_\beta\}$, $i.e.$ the set of the
  values of the function $d(x)$ on the nodes of $g$,
\item the \textit{operator} $\{O_{\alpha\beta}\}$, that is the set of
  values of the integral in the r.h.s. of
  Eq.~(\ref{eq:NumConvolution}) for all indices $\alpha$ and $\beta$,
  both running on the nodes of $g$.
\end{itemize}
Therefore, the primary functionality of the code is the construction
of these three objects. It is interesting to notice that
Eq.~(\ref{eq:NumConvolution}) reduces the Mellin convolution in
Eq.~(\ref{eq:MellinConvolution}) to a simple problem of linear algebra
and as such all the properties of linear transformations hold. More in
particular, Eq.~(\ref{eq:NumConvolution}) shows that $M_\alpha$
belongs to the category of distributions like $d_\beta$. In other
words, in this context a physical observable is also identified with a
distribution. In the next section I will show how the ingredients
listed above are sufficient for a number of interesting applications.

\paragraph{Applications.}

The original purpose of {\tt APFEL} was the solution of the
\textit{DGLAP evolution equations} in the factorisation scale $\mu$
that, in the framework discussed above, reduce to:
\begin{equation}\label{eq:DiscDGLAP}
\frac{{\rm d}{\bf f}_\alpha}{{\rm d}t} = {\bf \overline{P}}_{\alpha\beta}\cdot {\bf f}_\beta\,.
\end{equation}
with $t=\ln\mu^2$ and where ${\bf f}_\alpha$ is a vector of
distributions in the quark-flavour space, corresponding to the parton
distribution functions (PDFs) or to the fragmentation functions (FFs)
of the different partonic species, and
${\bf \overline{P}}_{\alpha\beta}$ is instead a matrix of operators in
the quark-flavour space, corresponding to the splitting functions. The
dot in the r.h.s. of Eq.~(\ref{eq:DiscDGLAP}) represents the product
in flavour space between the matrix ${\bf \overline{P}}$ and the
vector {\bf f} that gives rise to another
vector. Eq.~(\ref{eq:DiscDGLAP}) is a linear ordinary differential
equation in ${\bf f}_\alpha$ that can be solved using, for example,
the Runge-Kutta method.

\begin{figure}[h]
  \begin{center}
    \includegraphics[angle=270,width=0.8\textwidth]{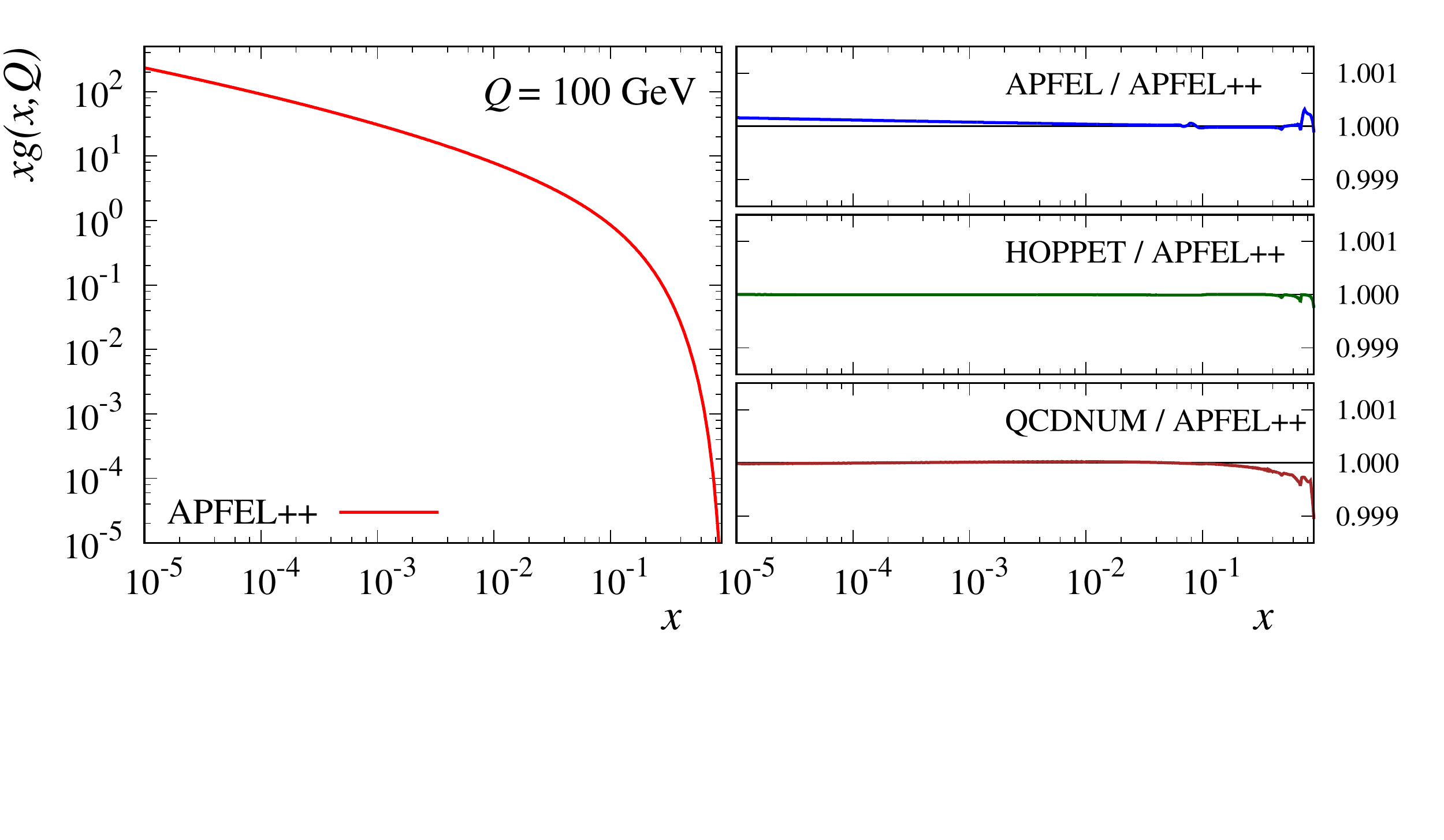}
  \end{center}
  \vspace{-60pt}
  \caption{\footnotesize The gluon PDF at 100 GeV. Left panel:
    distribution evolved in NNLO QCD with {\tt APFEL++}. Right panels:
    ratios to {\tt APFEL++} of the same distribution evolved with:
    {\tt APFEL} (upper plot), {\tt HOPPET}~\cite{Salam:2008qg}
    (central plot), and {\tt QCDNUM}~\cite{Botje:2010ay} (lower
    plot). The initial conditions for are those of
    Ref.~\cite{Dittmar:2005ed}.}
  \label{fig:dglap}
\end{figure}
As a practical application, Fig.~\ref{fig:dglap} displays the gluon
PDF evolved to $Q=100$ GeV at NNLO in QCD starting from the initial
conditions given in Ref.~\cite{Dittmar:2005ed}. The result obtained
with {\tt APFEL++} and shown in the left panel, is compared to the
three public codes: {\tt APFEL}, {\tt HOPPET}~\cite{Salam:2008qg}, and
{\tt QCDNUM}~\cite{Botje:2010ay}. The agreement between all these
codes is excellent with very small deviations mostly arising from the
different interpolation strategies.

The computation of the deep-inelastic-scattering (DIS) structure
functions in $ep$ collisions can also be addressed using the same
technology. In this case, the observable $F$ can be computes as:
\begin{equation}\label{eq:DiscSF}
F_\alpha = {\bf C}_{\alpha\beta}\cdot {\bf f}_\beta\,,
\end{equation}
where ${\bf f}_\alpha$ is a vector of distributions in flavour space,
corresponding to the relevant combinations of PDFs, and
${\bf C}_{\alpha\beta}$ is also a vector in flavour space but of
operators, corresponding to the coefficient functions. The dot in the
r.h.s now indicates the inner product in flavour space between
${\bf C}$ and ${\bf f}$ that gives rise to the scalar $F$.

\begin{figure}[h]
  \begin{center}
    \includegraphics[angle=270,width=0.8\textwidth]{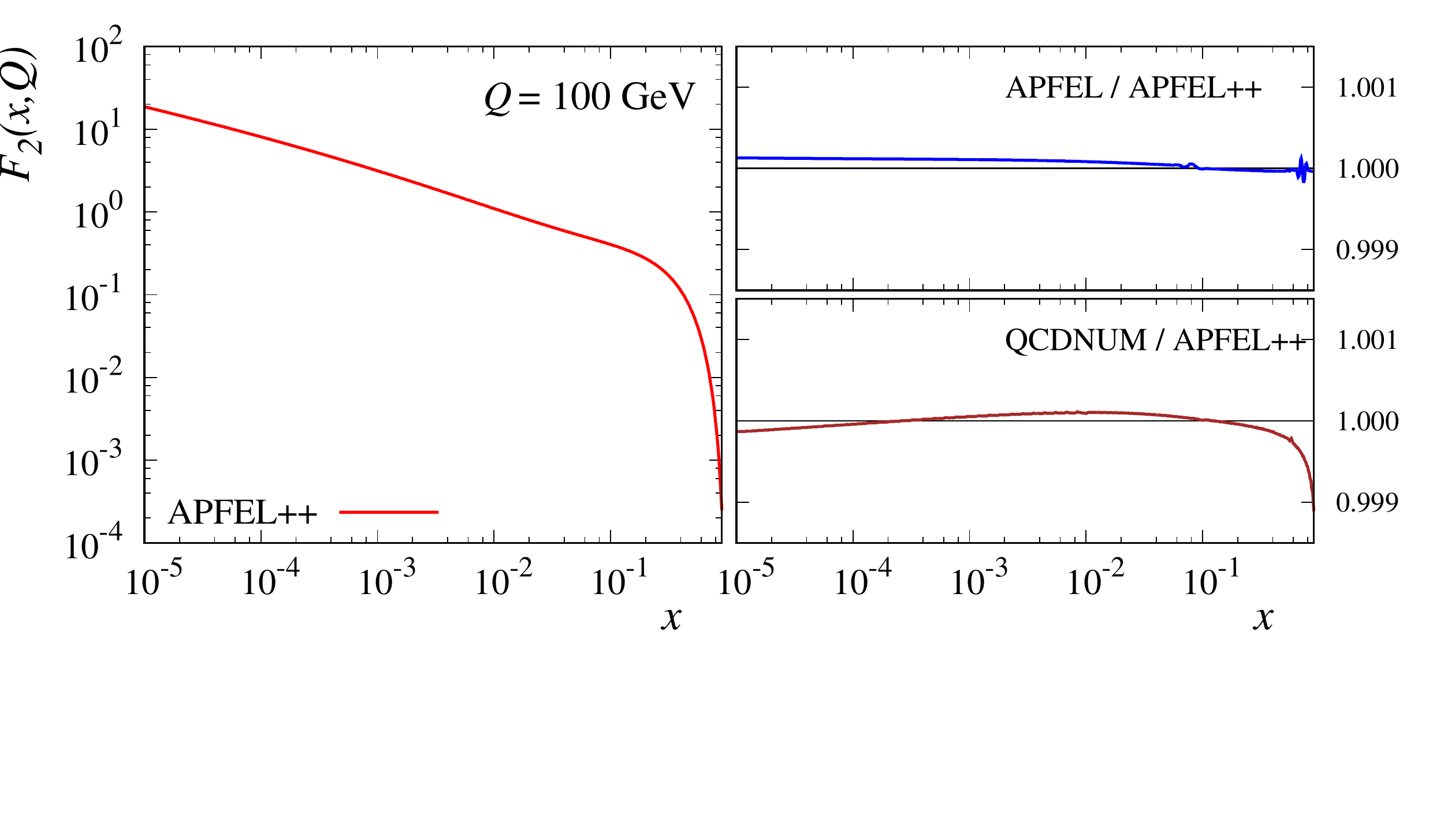}
  \end{center}
  \vspace{-60pt}
  \caption{\footnotesize The structure function $F_2$ at 100 GeV. Left
    panel: $F_2$ in NNLO QCD with {\tt APFEL++}. Right panels: ratios
    to {\tt APFEL++} of the same observable computed with: {\tt APFEL}
    (upper plot), and {\tt QCDNUM}~\cite{Botje:2010ay} (lower plot).}
  \label{fig:structurefunction}
\end{figure}
Fig.~\ref{fig:structurefunction} displays the prediction for the DIS
structure function $F_2$ at NNLO in QCD computed at $Q=100$ GeV. The
result obtained with {\tt APFEL++} is shown in the left panel and
compared to {\tt APFEL} and {\tt QCDNUM} in the right panels. The
agreement between these codes is again extremely good.

The method described above applies also to the convolution of two
operators, $O^{(1)}$ and $O^{(2)}$, that gives as a result to a third
operator $O^{(3)}$, as follows:
\begin{equation}
O^{(3)}_{\alpha\beta} =O^{(1)}_{\alpha\gamma}O^{(2)}_{\gamma\beta}=O^{(2)}_{\alpha\gamma}O^{(1)}_{\gamma\beta}\,,
\end{equation}
where the last equality follows from the definition of operator in
Eq.~(\ref{eq:NumConvolution}). This kind of products is relevant when
considering factorisation scale variations that typically generate
terms involving convolutions between hard cross sections and splitting
functions.

As an example, I have computed the convolution:
\begin{equation}\label{eq:operatorproduct}
  \left[P_{qq}^{(0)}(x) \otimes P_{qq}^{(0)}(x)\right] \otimes
  f(x)\,,\quad\mbox{with}\quad P_{qq}^{(0)}(x) =
  \left(\frac{1+x^2}{1-x}\right)_+\quad\mbox{and}\quad
  \underbrace{f(x)=1-x}_{\mbox{\small test function}}\,,
\end{equation}
both numerically, using the method discussed above, and analytically,
knowing that:
\begin{equation}\label{eq:P02an}
  \small
  P_{qq}^{(0)}(x) \otimes P_{qq}^{(0)}(x) =
  \left(\frac{4(x^2+1)\ln(1-x)+x^2+5}{1-x}\right)_+- \frac{(3x^2+1)\ln x}{1-x}
  - 4 +\left(\frac{9}{4}-\frac{2\pi^2}{3}\right)\delta(1-x)\,.
\end{equation}

\begin{figure}[h]
  \begin{center}
    \includegraphics[angle=270,width=0.8\textwidth]{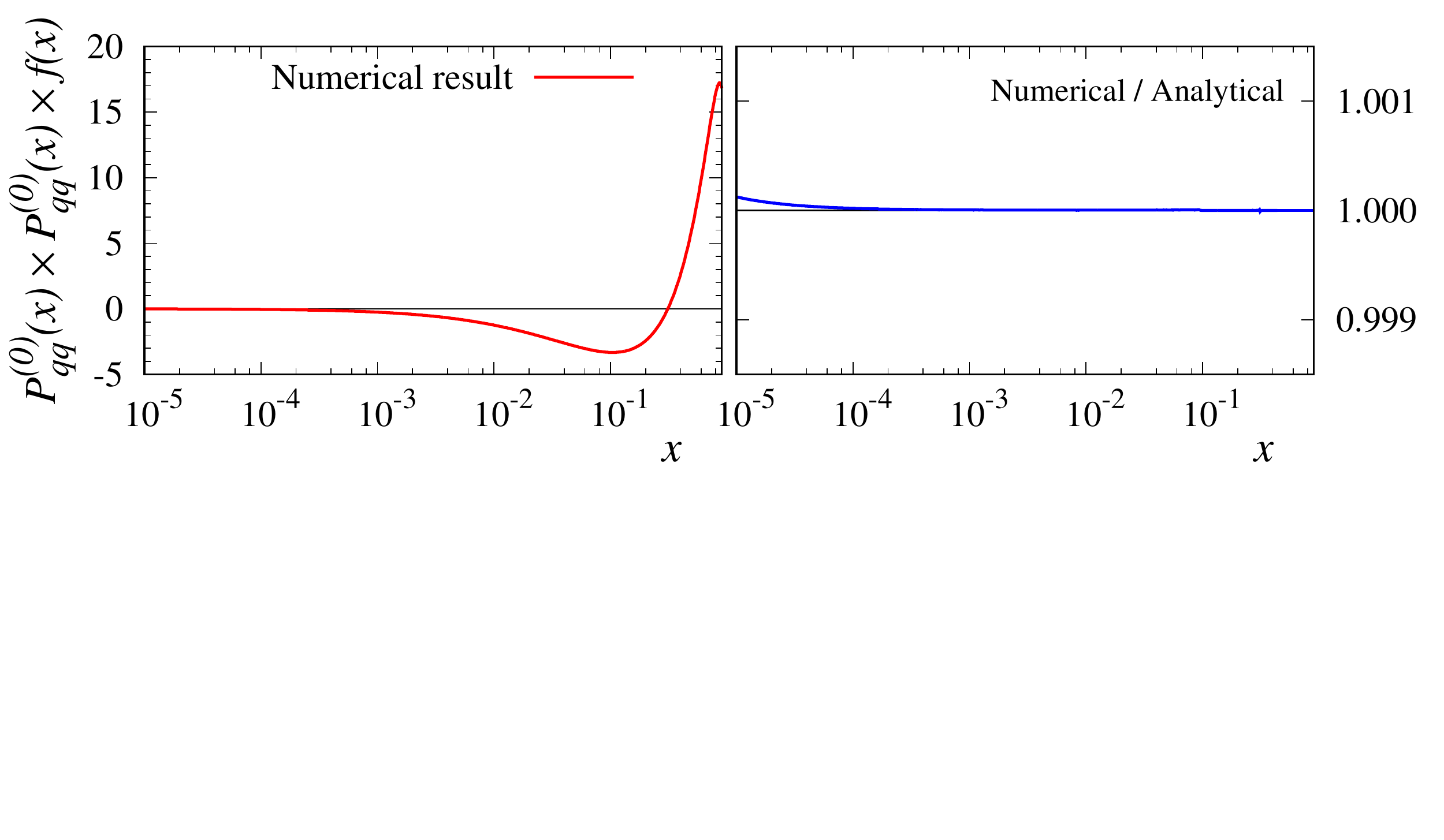}
  \end{center}
  \vspace{-100pt}
  \caption{\footnotesize Representation of
    Eq.~(\ref{eq:operatorproduct}). The left panel displays the result
    obtained computing the convolution
    $P_{qq}^{(0)}(x) \otimes P_{qq}^{(0)}(x)$ numerically while the
    right panel shows the ratio to the result obtained using the
    analytical expression in Eq.~(\ref{eq:P02an}).}
  \label{fig:operatorproduct}
\end{figure}

Fig.~\ref{fig:operatorproduct} shows an excellent agreement between
the numerical and the analytical computations, demonstrating that the
numerical convolution is very accurate over all the considered range.

Finally, this technology turns out to be useful also in the presence
of convolutions over more than one variable. Two interesting examples
are the computation of the Drell-Yan and of the semi-inclusive DIS
(SIDIS) cross sections. In this case, the typical integrals to compute
have the form:
\begin{equation}\label{eq:DoubleConv}
D(x,z) = \int_x^1\frac{d\xi}{\xi}\int_z^1\frac{d\zeta}{\zeta}\,O\left(\frac{x}{\xi},\frac{z}{\zeta}\right)\,d^{(1)}(\xi)\,d^{(2)}(\zeta)\,.
\end{equation}
Restricting ourselves to the computation of these cross sections to
$\mathcal{O}(\alpha_s)$, $i.e.$ next-to-leading order in QCD, a simple
inspection reveals that the expressions of the hard cross
sections~\cite{deFlorian:1997zj,Gehrmann:1997ez}, represented by the
``double'' function $O$ in Eq.~(\ref{eq:DoubleConv}), can be
decomposed as follows:
\begin{equation}
O(x,z) = \sum_i K_iC_i^{(1)}(x)C_i^{(2)}(z)\,,
\end{equation}
where $K_i$ are numerical factors, while $C_i^{(1)}$ and $C_i^{(2)}$
are ``single'' functions of the variables $x$ and $z$. This allows for
a simple extension of the procedure discussed above so that:
\begin{equation}
D_{\alpha\beta}\equiv D(x_\alpha,z_\delta) = \sum_i K_i
\left[O_{i,\alpha\gamma}^{(1)} d_\gamma^{(1)}\right]\left[O_{i,\beta\delta}^{(2)}d_\delta^{(2)}\right]\,,
\end{equation}
that is a bilinear combination of single operators. This combination
greatly reduces the complexity of the computation of the double
integral in Eq.~(\ref{eq:DoubleConv}) and thus makes it much faster.

\begin{figure}[h]
  \begin{center}
    \includegraphics[angle=270,width=1\textwidth]{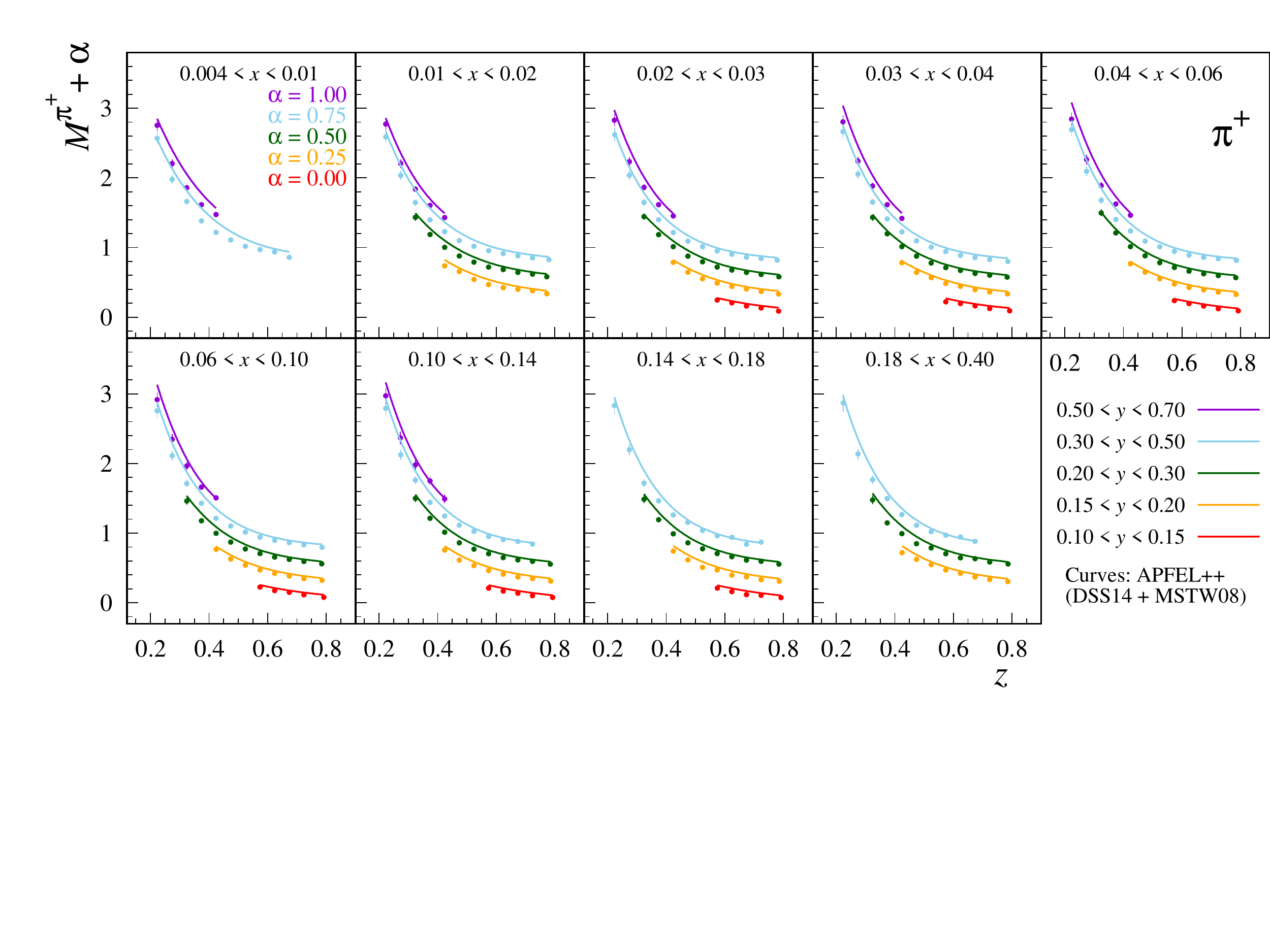}
  \end{center}
  \vspace{-100pt}
  \caption{\footnotesize Comparison between the SIDIS multiplicities
    for $\pi^+$ measured by COMPASS and the~\cite{Adolph:2016bga} and
    the NLO predictions computed with {\tt APFEL++} using the MSTW2008
    PDF set~\cite{Martin:2009iq} and the DSS14 FF
    set~\cite{deFlorian:2014xna}.}
  \label{fig:compass}
\end{figure}
I have applied this strategy to the computation of the SIDIS
multiplicities. In Fig.~\ref{fig:compass} the NLO predictions for
$\pi^+$ production obtained with {\tt APFEL++} using the MSTW2008 PDF
set~\cite{Martin:2009iq} and the DSS14 FF set~\cite{deFlorian:2014xna}
are compared to the COMPASS data~\cite{Adolph:2016bga}. Since these
measurements have been made public very recently, they have not been
included in any FF analysis yet. However, the predictions obtained
with {\tt APFEL++} are in fair agreement with the data.

\paragraph{Performance.}

The main foreseeable application of {\tt APFEL++} is in fits of PDFs
and/or FFs, where a large number of predictions have to be produced
very quickly in order to ensure that the fits converge in a reasonable
amount of time. Therefore, a lot of effort has been put to make the
computation of the evolution and of the observables as fast as
possible. To this purpose, {\tt APFEL++} implements a method to
tabulate operators and distributions over a grid in the scale $Q$ in
such a way that quantities that depend on both $x$ and $Q$, like
PDFs/FFs and structure functions, are obtained through a bidimensional
interpolation on the $(x,Q)$ grid. This is particularly useful in fits
of PDFs where thousands of predictions have to be computed at each
iteration of the fitting algorithm. Indicatively the tabulation with
{\tt APFEL++} of a structure function over an accurate $(x,Q)$ grid
takes a few hundredths of second. This has to be done once at every
iteration while the actual computation of the observables by
interpolation is extremely fast and, even for a very large number of
predictions, constitutes only a small fraction of the total computing
time.


\paragraph{Delivery.}

{\tt APFEL++} is publicly available from the website:
\begin{center}
  \url{https://github.com/vbertone/apfelxx}
\end{center}
However, the user should be aware that the code is still under
development and thus it may undergo substantial changes.

{\small \paragraph{Acknowledgements.} My thanks go to S.~Carrazza who
  is co-author of {\tt APFEL++}. I would also like to thank R. Sassot
  for providing me with the predictions of the SIDIS multiplicities
  for the COMPASS (and HERMES) data helping me debug the
  implementation in {\tt APFEL++}. I also thank D. Britzger for many
  useful discussions that have helped improve the code. Finally, I
  would like to thank N.~Hartland and J.~Rojo for a critical reading
  the manuscript. This work is supported by the European Research
  Council Starting Grant ``PDF4BSM''.}

\end{document}